\documentstyle{mn2e}
\begin{document}
\def\simlt{\mathrel{\rlap{\lower 3pt\hbox{$\sim$}}
        \raise 2.0pt\hbox{$<$}}}
\def\simgt{\mathrel{\rlap{\lower 3pt\hbox{$\sim$}}
        \raise 2.0pt\hbox{$>$}}}
\def\bj{b_{\rm\scriptscriptstyle J}}
\def\rt{r_{\rm\scriptscriptstyle T}}
\def\rp{r_{\rm\scriptscriptstyle P}}
\renewcommand{\labelenumi}{(\arabic{enumi})}

\title[A highly obscured and strongly clustered galaxy population]
{A highly obscured and strongly clustered galaxy population discovered with
the Spitzer Space Telescope}
\author[Manuela Magliocchetti et al.]
{\parbox[t]\textwidth{M. Magliocchetti$^{1,2}$, L. Silva$^1$, A.
Lapi$^2$, G. De Zotti$^{3,2}$, G.L. Granato$^{3,2}$, D. Fadda$^4$, L. Danese$^{2,3}$ }\\
 \\
{\tt $^1$ INAF, Osservatorio Astronomico di Trieste, Via Tiepolo 11, 34100,
Trieste, Italy}\\
{\tt $^2$ SISSA, Via Beirut 4, 34014, Trieste, Italy}\\
{\tt $^3$ INAF, Osservatorio Astronomico di Padova, Vicolo dell'Osservatorio
5, 35122 Padova, Italy}\\
{\tt $^4$ NASA Herschel Science Center, CalTech, MC 100-22, 770
South Wilson Avenue, Pasadena, CA 91125, USA}} \maketitle

\vspace{7cm}
\begin{abstract}
The $\sim 800$ optically unseen ($R>25.5$) 24$\,\mu$m-selected
sources in the complete {\it Spitzer} First Look Survey sample
(Fadda et al. 2006) with $F_{24\mu\rm m}\ge 0.35\,$mJy are found
to be very strongly clustered. If, as indicated by several lines
of circumstantial evidence, they are ultraluminous far-IR galaxies
at $z\simeq 1.6$--2.7, the amplitude of their spatial correlation
function is very high. The associated comoving clustering length
is estimated to be $r_0=14.0_{-2.4}^{+2.1}$ Mpc, value which puts
these sources amongst the most strongly clustered populations of
our known universe. Their $8\,\mu$m--$24\,\mu$m colours suggest
that the AGN contribution dominates above $F_{24\mu\rm m}\simeq
0.8\,$mJy, consistent with earlier analyses. The properties of
these objects (number counts, redshift distribution, clustering
amplitude) are fully consistent with those of proto-spheroidal
galaxies in the process of forming most of their stars and of
growing their active nucleus, as described by the Granato et al.
(2004) model. In particular, the inferred space density of such
galaxies at $z\simeq 2$ is much
higher than what expected from most semi-analytic models.\\
Matches of the observed projected correlation function $w(\theta)$
with models derived within the so-called Halo Occupation Scenario
show that these sources have to be hosted by haloes more massive
than $\simeq 10^{13.4} M_\odot$. This value is significantly
higher than that for the typical galactic haloes hosting massive
elliptical galaxies, suggesting a duration of the starburst phase
of massive high-redshift dusty galaxies of $T_B\sim 0.5$~Gyr.
\end{abstract}


\section{Introduction}

Understanding the assembly history of massive spheroidal galaxies
is a key issue for galaxy formation models.  The ``naive"
expectation from the canonical hierarchical merging scenario, that
proved to be remarkably successful in explaining many aspects of
large-scale structure formation, is that massive galaxies
generally form late and over a long period of time as the result
of many mergers of smaller haloes. On the other hand, there is
quite extensive evidence that massive galaxies may form at high
redshifts and on short timescales (see, e.g. Cimatti et al. 2004;
Fontana et al. 2004; Glazebrook et al. 2004; Giallongo et al.
2005; Treu et al. 2005; Saracco et al. 2006; Bundy et al. 2006),
while the sites of active star formation shift to lower mass
systems at later epochs, a pattern referred to as "downsizing"
(Cowie et al. 1996). In order to reconcile the observational evidence
that stellar populations in large
spheroidal galaxies are old and essentially coeval (Ellis et al.
1997; Holden et al. 2005) with the hierarchical
merging scenario, the possibility of mergers of evolved
sub-units (``dry mergers'') has been introduced (van Dokkum et al.
2005; Faber et al. 2006; Naab et al. 2006). This mechanism is
however strongly disfavoured by studies on the evolution of the
stellar mass function (Bundy et al. 2006).

Key information, complementary to optical/IR data, has come from
sub-millimeter surveys (Hughes et al. 1998; Eales et al. 2000;
Knudsen et al. 2006) which have found a large population of
luminous sources at substantial redshifts (Chapman et al. 2005).
However, the interpretation of this class of objects is still
controversial (e.g. Granato et al.
2004; Kaviani et al. 2003; Baugh et al. 2005).

The heart of the problem are the masses of the objects: a large
fraction of present day massive galaxies already assembled at
$z\sim 2-3$ would be extremely challenging for the standard view
of a merging-driven  growth.   Measurements of clustering
amplitudes are a unique tool to estimate halo masses at high $z$,
but complete samples comprising at least several hundreds of
sources are necessary. This is far more than what detected by
sub-mm surveys, that have therefore only provided tentative
clustering estimates (Blain et al. 2005).

Here we report evidence of strong clustering for the optically very
faint ($R>25.5$) sources included in the complete $24\, \mu$m sample
obtained from the Spitzer first cosmological survey (Fadda et al.
2006). Comparisons with template spectral energy distributions and
up-to-date models for galaxy formation and evolution set these
objects at $z \sim 2$. The clustering properties and the counts of
such sources are consistent with them being very massive
proto-spheroidal galaxies in the process of forming most of their
stars. Their comoving number density appears to be much higher
than what expected from most semi-analytic models.

The layout of the paper is as follows. In \S\,2 we describe the
sample selection. In \S\,3 we investigate the source photometric
and spectroscopic properties. In \S\,4 we derive the two point
angular correlation function, while in \S\,5 we present its
implications for source properties, and in particular for their
halo masses, in the light of the so-called Halo Occupation Model.
Comparisons with model predictions are dealt with in \S\,6. Our
main conclusions are summarized in \S\,7.

Throughout this work we adopt a flat cosmology with a matter
density $\Omega_m=0.3$ and a vacuum energy density
$\Omega_\Lambda=0.7$, a present-day value of the Hubble parameter
in units of $100$ km/s/Mpc $h=0.7$, and rms density fluctuations
within a sphere of $8 h^{-1}$ Mpc radius $\sigma_8=0.8$ (Spergel
et al. 2003).

\section{The sample selection}
\subsection{The Parent Catalogue}

Our analysis is based on the 24$\, \mu$m data obtained during the
first cosmological survey performed by the Spitzer Space Telescope
(First Look Survey). Observations and data reduction are
extensively described in Fadda et al. (2006). Briefly, the survey
consists of a shallow observation of a $2.5^\circ\times 2^\circ$
area centered at $(17^h18^m, +59^\circ30^\prime)$ (main survey)
and of a deeper observation on a smaller region of the sky
(verification survey) overlapping with the first one.

Observations were performed using the MIPS (Multi Imaging
Photometer for Spitzer, Rieke et al. 2004), whose spatial
resolution at 24$\, \mu$m is 5.9$^{\prime\prime}$ FWHM. Approximately
$\sim 17000$ sources have been extracted with
signal-to-noise-ratio (SNR) greater than five down to $\sim
0.2$~mJy in the main survey and to $\sim 0.1$~mJy in the
verification survey. Astrometric errors depend on the SNR, varying
between $0.35^{\prime\prime}$ and $1.1^{\prime\prime}$ for sources
detected at 20--5$\sigma$levels.
The main survey is estimated to be $> 90$\% complete down to a
limiting flux $F_{24\mu\rm m}=0.35$~mJy (Fadda et al. 2006).

Optical counterparts have been obtained by Fadda et al. (2006) for
most of the 24$\,\mu$m sources by cross-correlating galaxies in
the MIPS catalogue with the $R$-band KPNO observations of Fadda et
al. (2004) and -- for objects with $R\le 18$ -- with sources from
the Sloan Digital Sky Survey (Hogg et al., in preparation). These
two optical data-sets cover in a roughly homogeneous way most of
the area probed by the 24$\, \mu$m main survey, except for three
corners. The typical limiting magnitude of the joint SDSS+KPNO
observations is $R=25.5$, and $\sim 82\%$ of the 24$\, \mu$m
of all sources in the MIPS survey 
are reported to have an optical counterpart brighter than
this limit.

Despite ongoing efforts (Marleau et al. 2003; Choi et al. 2006;
Yan et al. 2005), there is still no homogeneous redshift
information on the sources making up the MIPS 24$\, \mu$m catalogue,
{except for a very small area overlapping with the GOODS/CDFS field
(Caputi et al. 2006)}.
However, redshift estimates can be obtained from photometric data,
taking advantage of the Spitzer Infrared Array Camera (IRAC)
survey which covers an extensive portion of the MIPS field
(Lacy et al. 2005).

The main IRAC survey has covered an area of 3.8 square degrees in
the four channels centered at 3.6, 4.5, 5.8 and 8$\, \mu$m, reaching
a $\sim 100$\% completeness level respectively at $\sim 40$, $\sim 40$, $\sim
100$ and $\sim 100~\mu$Jy (Fig. 3 of Lacy et al.
2005). The positional accuracy goes from $\sim
0.25^{\prime\prime}$ for high signal-to-noise sources to
1$^{\prime\prime}$ at the lowest flux levels.

Investigations of the Spectral Energy Distributions (SED) of
prototype sources such as M~82, Arp~220, and Mkn~231 (the latter
with mid-IR luminosity probably powered by the presence of an AGN)
shows that the tightest constraints on the redshifts of very
distant galaxies with intense star-formation come from the 8$\,
\mu$m IRAC channel, since such sources are expected to be very
weak at shorter wavelengths (see also Yan et al. 2005). Therefore,
in the following we will only consider data from the 8$\,\mu$m
channel.

\subsection{Matching procedure}

\begin{figure}
\vspace{8cm}  
\includegraphics{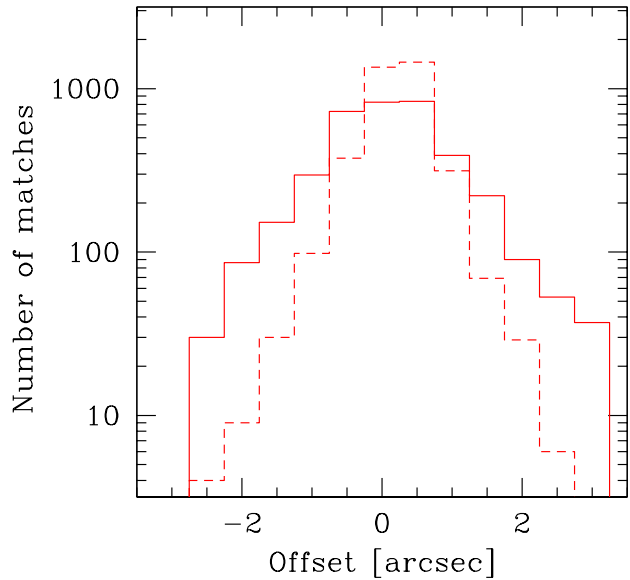} \caption{Distribution of the residuals $\Delta
x=$RA$_{24\mu\rm m}-$RA$_{8\mu\rm m}$ (solid line) and $\Delta y$=Dec$_{24\mu\rm
m}-$Dec$_{8\mu\rm m}$ (dashed line) between 24$\, \mu$m and 8$\, \mu$m positions.
\label{fig:residuals}}
\end{figure}

We have looked for the 8$\,\mu$m counterparts to MIPS sources over
the 2.85 square degrees region $257.7^\circ \simlt$ RA(2000) $\simlt
261^\circ$ and 58.6$^\circ \simlt$ Dec $\simlt 60.3^\circ$,
for which both 8$\, \mu$m and $R$-band observations are available.
In this area there are 7592 24$\, \mu$m sources
and 8,646 8$\, \mu$m sources above the respective completeness limits
of 0.35~mJy and 0.1~mJy.

We identified as the counterpart to a MIPS source the IRAC source
with a positional separation less than a suitably chosen radius.
As mentioned in \S\,2.1, the positional accuracies for both MIPS
and IRAC sources with $\hbox{SNR}=5$ is $\simeq 1''$, so that we
expect a rms positional difference due to astrometric errors of
$\sim \sqrt{2''} \simeq 1.4''$. Moreover, the IRAC astrometry is
based on 2MASS, while that of the 24$\, \mu$m sources is related
to SDSS. Although both systems are very accurate, a small
systematic offset may be present.

Tackling the problem from a more pragmatic point of view, we
considered the distribution of the residuals $\Delta x={\rm
RA}_{24\mu\rm m}-{\rm RA}_{8\mu\rm m} $, $\Delta y={\rm
Dec}_{24\mu\rm m}-{\rm Dec}_{8\mu\rm m}$ between the positions of
all 24$\, \mu$m and 8$\, \mu$m pairs with separations $|\Delta x|$
and $|\Delta y|$ less than $5''$. The distribution of residuals shows a
strong concentration of points near $\Delta x\simeq -0.14$ and
$\Delta y \simeq 2.24 \cdot 10^{-3}$ arcsec, values which can be
taken as the mean positional offsets between the 24$\, \mu$m and
8$\, \mu$m reference frames.

We have corrected for this effect and {in
Fig.~\ref{fig:residuals} we plot the histogram of the number of
matches as a function of $\Delta x$ (solid line) and $\Delta y$
(dashed line) offsets. The distributions} now correctly peaks near
zero offset with a rms value of about $1''.5$, in agreement with
the above simple estimate. We have then chosen a $3''$ matching
radius -- equivalent to about $2\sigma$ -- which should therefore
include $\sim 95\% $ of the true identifications. The probability
that an 8$\, \mu$m source falls by chance within the search radius
from a 24$\, \mu$m source (equal to the surface density of 8$\mu$m
sources times the area within such radius) is $\simeq 6.6\times
10^{-3}$. Increasing the matching radius increases the number of
interlopers more than that of true counterparts.

The above procedure yielded 3429 $F_{24\mu\rm m}\ge 0.35$~mJy
24$\, \mu$m sources endowed with an 8$\, \mu$m counterpart. Given
the completeness limit of the IRAC survey, we will then assume the
remaining MIPS objects to have an 8$\, \mu$m counterpart fainter
than 0.1 mJy.

\begin{figure*}
\begin{center}
\vspace{8cm}  
\includegraphics{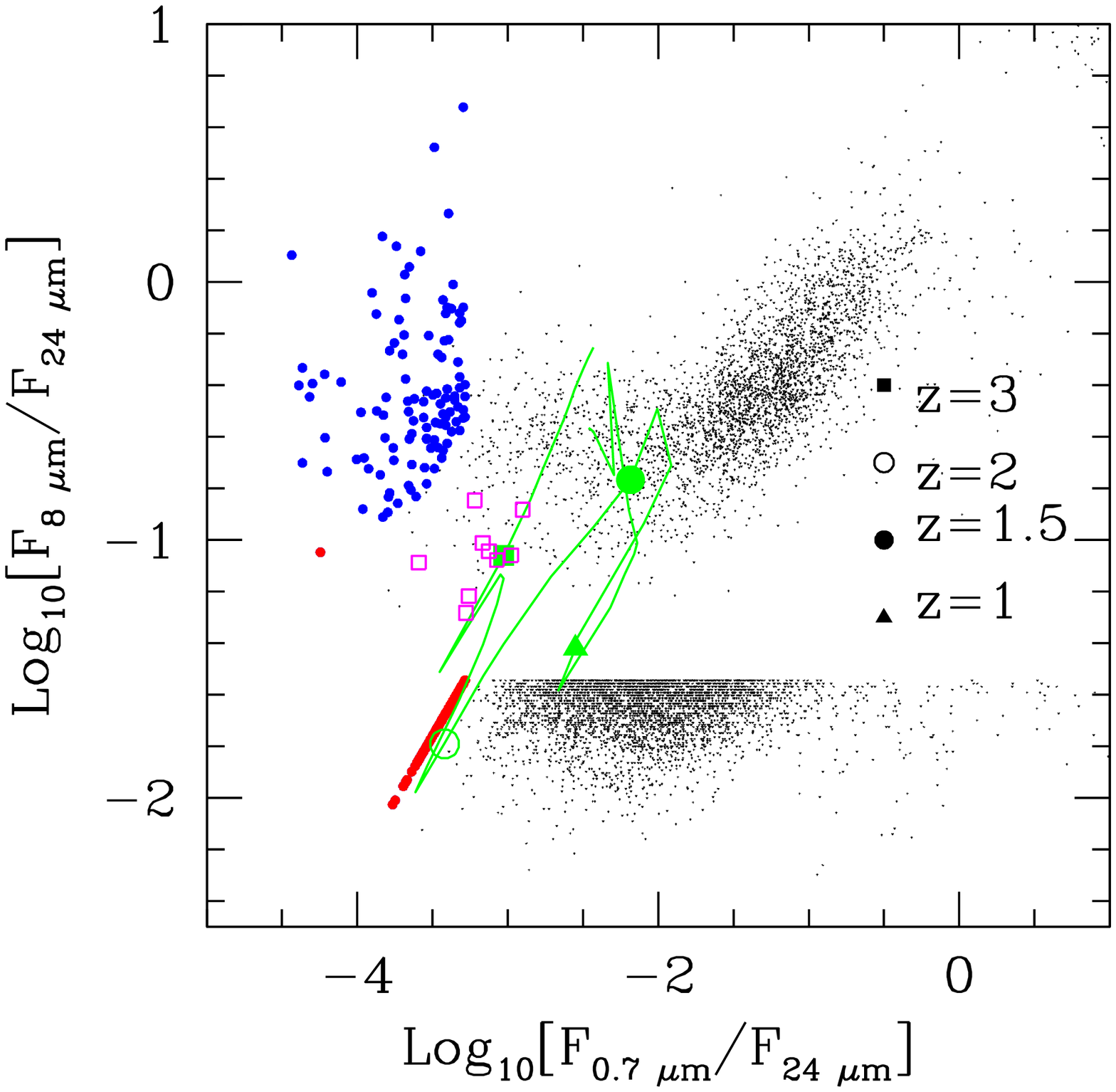} \includegraphics{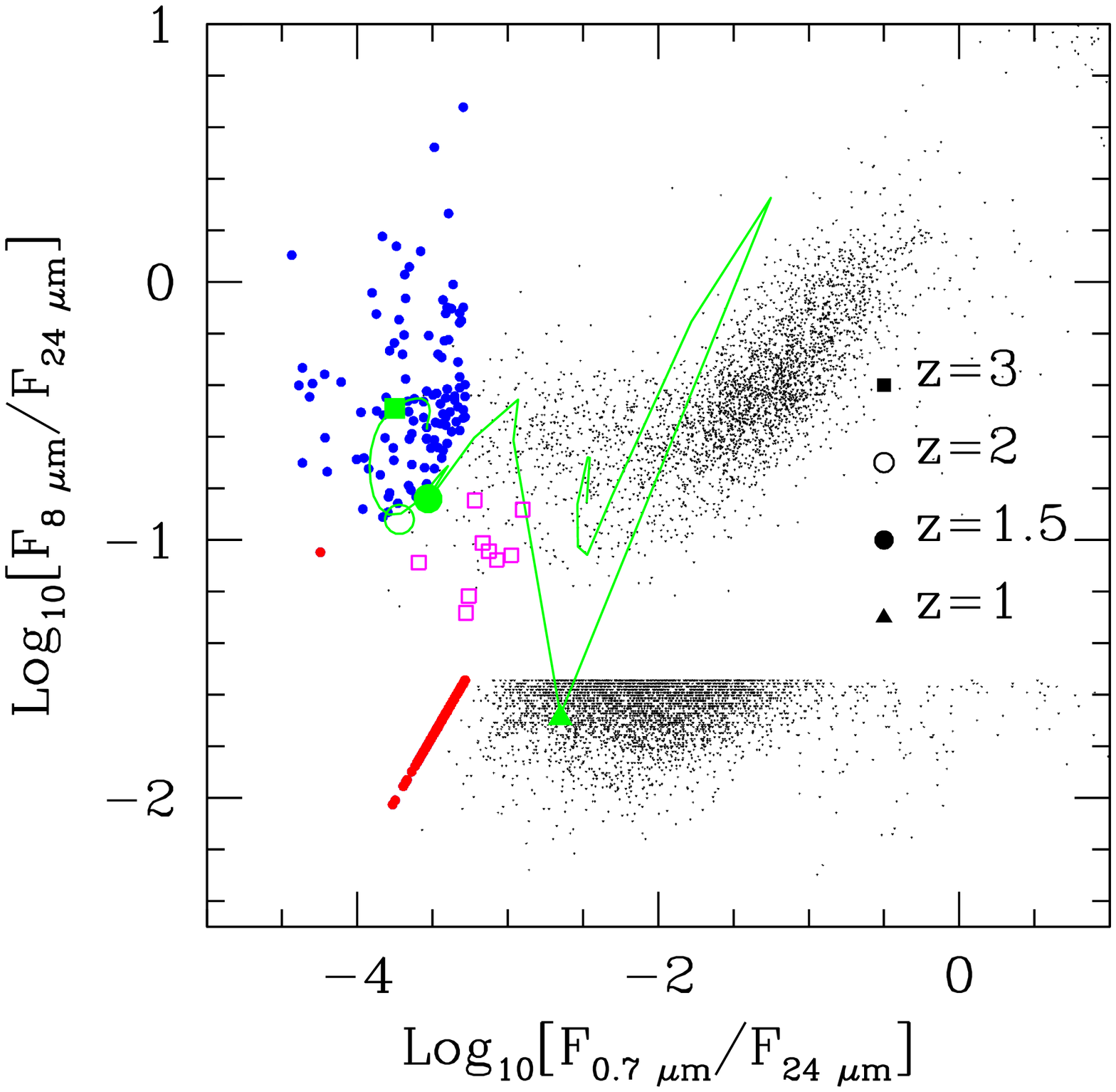}
\caption{Distribution of the 8$\, \mu$m to 24$\, \mu$m vs the
0.7$\, \mu$m to 24$\, \mu$m flux ratios for the 7592 $F_{24\mu \rm
m}\ge 0.35$~mJy MIPS sources in the $2.85\,\hbox{deg}^2$ region
covered by both the KPNO and the IRAC surveys. These are compared
with the computed colour-colour tracks as a function of redshift
(green solid lines) for the Arp~220 (left-hand panel) and the
Mkn~231 (right-hand panel) SEDs. For sake of clarity, objects with
an 8$\, \mu$m counterpart fainter than 0.1 mJy have been given
$F_{8\mu\rm m}=10^{-2}$~mJy and are responsible for the apparent
gap observed in the distribution along the $y$ axis, while sources
without an optical counterpart in the KPNO catalogue have been
attributed $R=25.5$ and are represented by the filled circles.
{Blue circles indicate objects with
 $F_{8\mu\rm m}/F_{24\mu\rm m}> 0.1$, while the red ones are for those with
$F_{8\mu\rm m}/F_{24\mu\rm m}< 0.1$. Note that, although this in
not clear from the figure, because many red points are piling up
in the same spot, objects with $F_{8\mu\rm m}/F_{24\mu\rm m}< 0.1$
are much more numerous than those with $F_{8\mu\rm m}/F_{24\mu\rm
m}> 0.1$.} Open squares represent sources from the Pope et al.
(2006) sample with $F_{24\mu\rm m}\ge 0.15$~mJy.
\label{fig:colours}}
\end{center}
\end{figure*}

\begin{figure*}
\vspace{8cm}  
\includegraphics{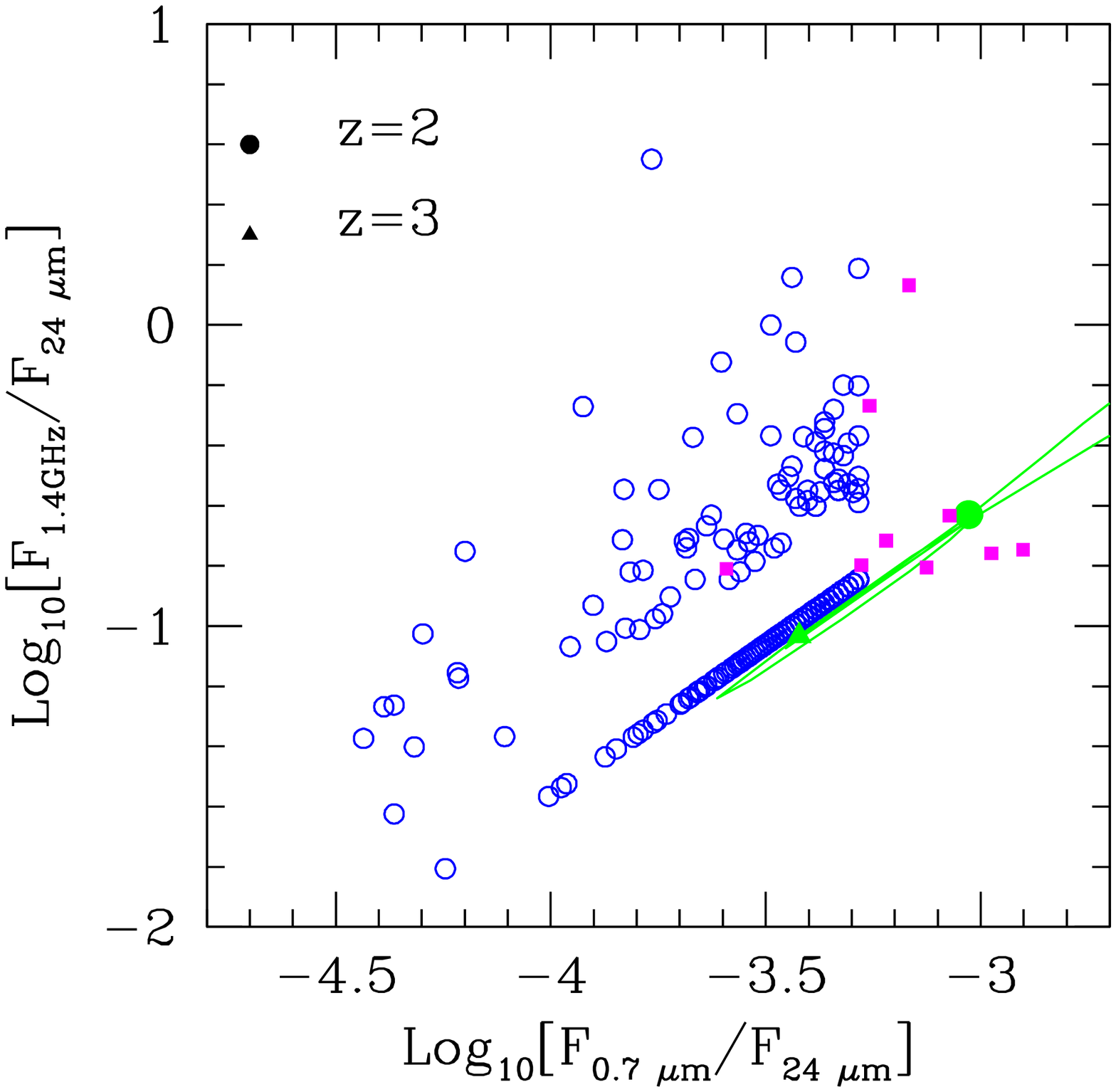} \includegraphics{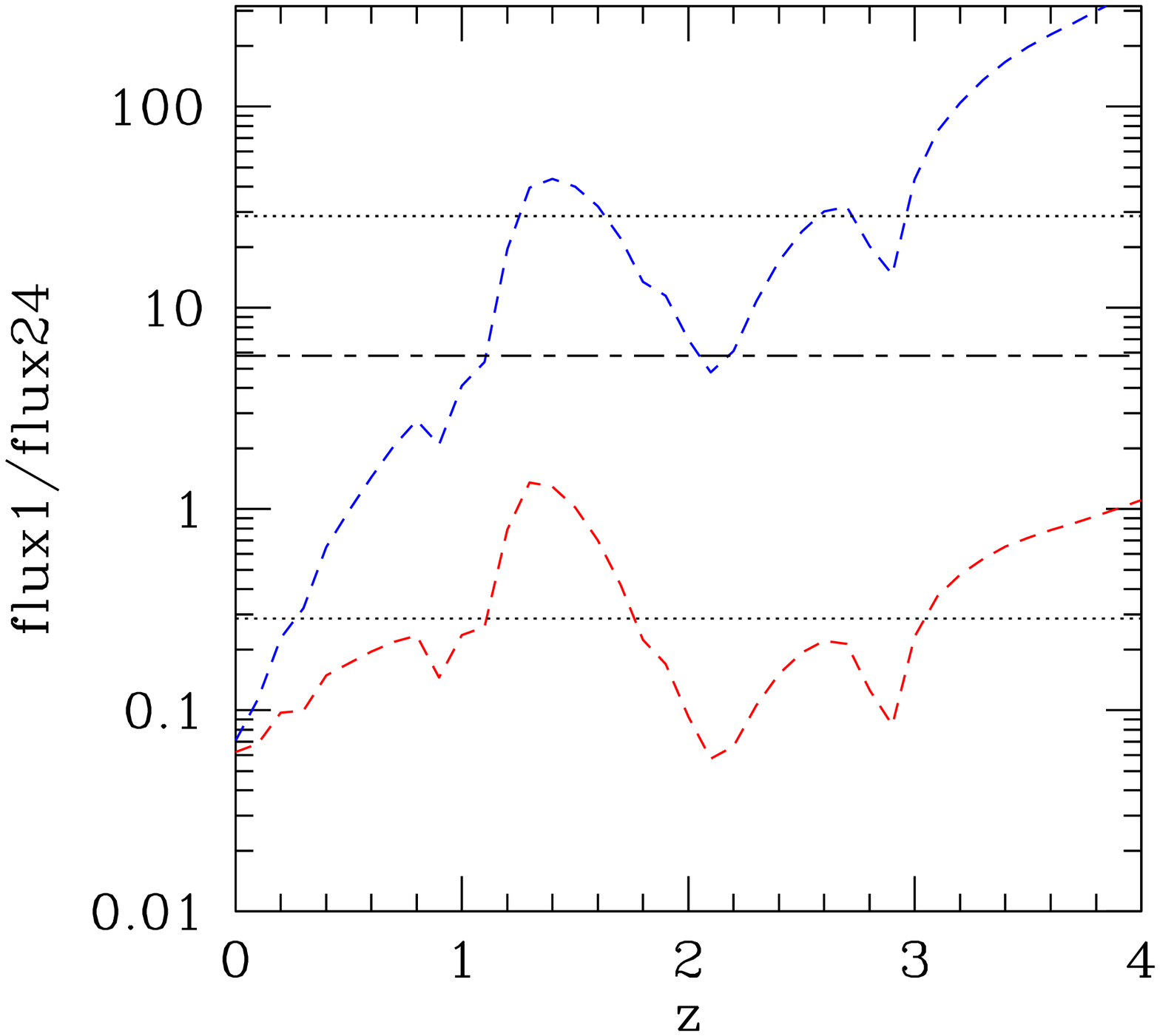}
\caption{{\it Left-hand panel}: distribution of the 1.4 GHz to
$24\, \mu{\rm m}$ vs the $0.7\, \mu{\rm m}$ to $24\, \mu{\rm m}$
flux ratios for the 793 $F_{24\mu \rm m}\ge 0.35$~mJy, $R> 25.5$
MIPS sources present in the $3.97\,\hbox{deg}^2$ region considered
in this work. For sake of clarity, objects fainter than the 0.1
mJy detection limit at 1.4 GHz have been attributed $F_{1.4\rm
GHz} = 5\cdot 10^{-2}\,$~mJy and produce the diagonal line on the
bottom right corner of the plot. Sources undetected by the KPNO
survey have been assigned $R = 25.5$. The filled squares indicate
objects from the Pope et al. (2006) sample with $F_{24\mu\rm m}
\ge 0.15\,$mJy, while the solid (green) line shows the track, as a
function of redshift, yielded by the Arp~220 SED. {\it Right-hand
panel}: redshift dependence of the $F_{850\mu\rm m}/F_{24\mu\rm
m}$ (upper curve) and $F_{1.4\rm GHz}/F_{24\mu\rm m}$ (lower
curve) flux density ratios for the Arp~220 SED. The upper dotted
line shows the ratio between the limiting fluxes of our sample
($F_{24\mu\rm m}=0.35$~mJy) and of the dataset by Sawicki \& Webb
(2005; $F_{850\mu\rm m}=10$~mJy), while the short-long dashed one
corresponds to decreasing $F_{850\mu\rm m}$ to 2 mJy. The lower
dotted line is the ratio between the limiting 24$\mu$m and 1.4 GHz
fluxes ($F_{1.4\rm GHz}=0.1$) for our sample.
\label{fig:radioarp}}
\end{figure*}

\section{Photometric and spectroscopic properties}\label{photspec}

The distribution of 8$\, \mu$m to 24$\, \mu$m vs  0.7$\, \mu$m to
24$\, \mu$m flux ratios for all the 7592 $F_{24\mu \rm m}\ge
0.35$~mJy MIPS sources in the $2.85\,\hbox{deg}^2$ region covered
by both the KPNO and the IRAC surveys is reported in
Fig.~\ref{fig:colours}. $R$ magnitudes have been converted to
0.7$\mu$m fluxes using the calibration in Fukugita, Shimasaku \&
Ichikawa (1995). For sake of clarity, objects with an 8$\, \mu$m
counterpart fainter than 0.1 mJy have been given $F_{8\mu\rm
m}=10^{-2}$~mJy and are responsible for the apparent gap observed
in the lower part of the $F_{8\mu\rm m}/F_{24\mu\rm m}$ axis,
while sources without an optical counterpart in the KPNO catalogue
have all been given R=25.5 and are represented by {the red or
blue filled circles. Blue circles indicate objects with
 $F_{8\mu\rm m}/F_{24\mu\rm m}> 0.1$, while the red ones are for those with
$F_{8\mu\rm m}/F_{24\mu\rm m}< 0.1$}.

The solid (green) lines show, as a function of redshift, the
colours corresponding to the SEDs of Arp~220 (left-hand panel), {a
well studied local starburst galaxy -- featuring in its mid-IR spectrum signatures of heavy dust
absorption (Spoon et al. 2004) -- found to provide to a first approximation
a good template to describe the energy output
of high-redshift galaxies undergoing intense star-formation (see e.g. Pope et al. 2006)},
and of Mkn 231 (right-hand panel), a
prototype absorbed AGN dominating the mid-IR emission, hosted in a
galaxy with very intense star formation. This figure shows that,
for both source types, extreme 24$\mu$m to $R$-band flux ratios
(or $R> 25.5$) likely correspond to sources at $z\simeq 1.6$--3.

This conclusion is supported by the comparison of the distribution
of the $F_{1.4\rm GHz}/F_{24\mu\rm m}$ vs $F_{0.7\mu\rm
m}/F_{24\mu\rm m}$ colours for the $R>25.5$ sources from the
complete MIPS sample with the track, as a function of redshift,
yielded by the Arp~220 SED (left-hand panel of
Fig.~\ref{fig:radioarp}). Radio data come from the $20$~cm radio
survey performed by Condon et al. (2003) on 82\% of the 24$\,
\mu$m field down to a limiting flux of 0.1~mJy. Only 86 out of the
793 $R>25.5$, $F_{24\mu \rm m}\ge 0.35$~mJy sources (11\% of the
sample) are detected. The lower dashed curve in the right-hand
panel of Fig.~\ref{fig:radioarp} details the redshift dependence
of the $F_{1.4\rm GHz}/F_{24\mu\rm m}$ ratio for the Arp~220 SED,
showing that sources undergoing intense star-formation and endowed
with 24$\, \mu$m fluxes close to the MIPS detection limit, present
1.4 GHz fluxes below the 0.1~mJy threshold of Condon's survey {
only if $z<1.2$ or $1.7 \le z \le 3$. Combining this latter piece
of information with the trend of the $F_{1.4\rm GHz}/F_{24\mu\rm
m}$ vs $F_{0.7\mu\rm m}/F_{24\mu\rm m}$ distribution, we can
conclude that most of the $R>25.5$ MIPS sources are consistent
with them being starburst galaxies placed in the redshift range
$1.6 \simlt z \simlt 3$. Although the above arguments do not exclude the
possibility that some of our sources are actually at $z\simeq 1$,
a really extreme extinction would be required to make them fainter
than $R=25.5$, and therefore such sources must be very rare. This
is directly confirmed by the spectroscopic observations summarized
below, that did not find $R\ge 25.5$ objects at $z<1.7$.}

The upper dashed curve in the right-hand panel of Fig.~\ref{fig:radioarp}
shows that galaxies with a SED similar to Arp 220, fluxes
$F_{24\mu\rm m}\simeq 0.35\,$mJy,
and lying in the redshift range $1.6 \le z \le 3$ have
$F_{850\mu\rm m}\simlt 12$~mJy. As pointed out by Houck et al.
(2005), sources with higher AGN contribution (in general brighter
at 24$\,\mu\rm m$ than the above limit) which therefore present
hotter dust, exhibit $F_{850\mu\rm m}/F_{24\mu\rm m}$ ratios lower
by a typical factor of 3--5 than those of (sub)-mm selected
sources {($F_{850\mu\rm m}/F_{24\mu\rm m}\simeq 5$; see e.g. Lutz et al. 2005)},
generally high-$z$ starburst galaxies (Ivison et al.
2004; Egami et al. 2004; Frayer et al. 2004; Charmandaris et al.
2004; Pope et al. 2006). A small fraction (151
arcmin$^{2}$) of the area covered by the First Look Survey has
been observed by Sawicki \& Webb (2005) with SCUBA on JCMT. These
authors report the detection ($S/N > 3.5$) of ten sources with
$F_{850\mu{\rm m}}\ge 10$~mJy. As expected on the basis of the
above discussion, none of them belongs to the complete sample of
$R>25.5$ MIPS sources although one of the detected objects is just
below the 0.35~mJy limit (J171736.9+593354, $F_{24\mu\rm m}=
0.32$~mJy).

Thus, radio, sub-mm, mid-IR and optical photometric data converge
in indicating that the process of extracting optically faint
sources from samples selected at 24$\, \mu\rm m$ singles out
star-forming galaxies in the redshift range $1.6\simlt z \simlt
3$. As pointed out by Houck et al. (2005) such a redshift range is
determined by well established selection effects. The requirement
for the sources to be optically very faint ($R > 25.5$) forces
them to $z \ge 1$, because obscuration is higher in the rest-frame
UV. On the other hand, the deep and broad $9.7\,\mu$m silicate
absorption feature, which is common in ultra-luminous infrared
galaxies (Armus et al. 2004; Higdon et al. 2004; Spoon et al.
2004) works against inclusion in $24\,\mu$m-selected samples of
objects in the redshift range $1 \simlt z \simlt 1.6$, while the
strongest PAH emission feature (set in the rest-frame at $\lambda
= 7.7\mu$m) enters the $24\,\mu$m filter at $z\sim 2.1$, and
another relatively strong PAH line ($\lambda = 6.2\mu$m) appears
at 24$\, \mu\rm m$ for $z\sim 2.9$. Ultra-luminous IR galaxies at
still higher redshifts are expected to become increasingly rare
because of the dearth of very massive galactic haloes.\\

The above conclusion is fully borne out by spectroscopic data,
although only obtained for a limited number of sources. Yan et al.
(2005) obtained low-resolution spectra with the Spitzer InfraRed
Spectrograph (IRS) for eight First Look Survey sources with
24$\mu$m fluxes brighter than 0.9~mJy. Further colour constraints
which were applied include: $\log_{10}(\nu F_\nu(24 \rm \mu m)/\nu
F_\nu(8 \rm \mu m))\ge 0.5$ and $\log_{10}(\nu F_\nu(24 \rm \mu
m)/\nu F_\nu(0.7 \rm \mu m))\ge 1.0$. Three of these sources
(namely IRS2, IRS8, and IRS9) have $R >25.5$. All three lie in the
redshift range $1.8 \simlt z \simlt 2.6$ ($z_{\rm IRS 2}=2.34$;
$z_{\rm IRS 8}=2.6$; $z_{\rm IRS 9}=1.8$). IRS2 and IRS9 show
strong PAH emission lines and moderate silicate absorption in
their spectra, while IRS8 only presents strong silicate
absorption. IRS9 has also been observed with MIPS at 70$\mu$m and
with MAMBO at 1.2~mm, and found to have fluxes of respectively
42~mJy and 2.5~mJy. The estimated bolometric luminosities (Yan et
al. 2005) are $L_{\rm bol}=1.83 \cdot 10^{13} L_\odot$ (IRS9),
$L_{\rm bol}=4.3 \cdot 10^{13} L_\odot$ (IRS2), and $L_{\rm bol}=2
\cdot 10^{13} L_\odot$ (IRS8).

Houck et al. (2005) used the Spitzer Telescope to image at 24$\,
\mu$m a $9\,\hbox{deg}^2$ field within the NOAO Deep Wide-Field
Survey region down to a flux of 0.3~mJy. Thirty-one sources, with
$F_{24\mu\rm m}\ge 0.75$~mJy and $R \ge 24.5$ have further been
observed with the IRS. Redshift determinations were possible for
17 of them, including 13 sources with $R > 25.5$. Again, the
measured redshifts are all in the range $1.7 < z < 2.6$, except
possibly for one object which appears to have $z=0.7$, but this is
the least secure determination due to a poor spectrum beyond
30$\mu$m.

Eighteen optically faint ($R > 23.9$) sources from the Spitzer
First Look Survey, with $F_{24\mu\rm m}> 1$~mJy and 20 cm
detections to a limit of $115\,\mu$Jy, have been observed with the
IRS by Weedman et al. (2006). All sources with $R > 24$ lie in the
range $1.7 < z < 2.5$.

Furthermore, Pope et al. (2006) have recently
presented 24$\, \mu$m observations of 35 sub-mm
selected sources with 850$\mu$m fluxes $\simgt 2$~mJy.  Nine of
these sources have 24$\, \mu$m fluxes brighter than 0.15~mJy, the
limit of the First Look Survey in the {\it verification} region,
and $i_{\rm AB} \ge 24$ (7 have $i_{\rm AB} \ge 26.2$). All of
them have spectroscopic or photometric redshifts in the range
$z\sim [1.7-2.7]$. Their colours, shown in Fig.~\ref{fig:colours}
by the magenta open squares and in the left-hand panel of
Fig.~\ref{fig:radioarp} by filled squares, lie in those regions
identified by the $F_{24\mu\rm m}\ge 0.35$~mJy and $R>25.5$ MIPS
objects, confirming that the above selection singles out galaxies
with properties similar to those detected by (sub)-mm surveys.

{On the basis of the above results obtained on both
photometric and spectroscopic grounds, we can then confidently
assume that sources with $F_{24\mu\rm m}\ge
0.35$~mJy and $R>25.5$ typically identify
star-forming galaxies in the redshift interval $1.6\simlt z\simlt
3$. Such a conclusion will be further strengthened in \S\,6, 
in the light of the most
up-to-date models for galaxy formation and evolution.}

\subsection{Starburst vs AGN components}
There are 510 sources with $R>25.5$ and $F_{24\mu\rm m}\ge
0.35$~mJy in the $2.85\,\hbox{deg}^2$ region where both 24~$\mu$m
and 8~$\mu$m data are available (see \S\,2.2).
Figure~\ref{fig:colours} illustrates that, for $z\simeq 1.5$--3,
``AGN-dominated'' and ``starburst-dominated'' SEDs have different
$F_{8\mu\rm m}/F_{24\mu\rm m}$ ratios, the dividing line being set
at $F_{8\mu\rm m}/F_{24\mu\rm m}\simeq 0.1$ (see also Yan et al.
2005): starburst galaxies (Arp~220-like SED) are generally below
this value, while AGN-powered sources (Mkn~231-like SED) are above
it. Most (401) of the 510 sources within the overlapping IRAC-MIPS
area present $F_{8\mu\rm m}/F_{24\mu\rm m} < 0.1$ so that -- to a
first order -- can be classified as ``starbursts''.

\begin{figure}
\begin{center}
\vspace{8cm}  
\includegraphics{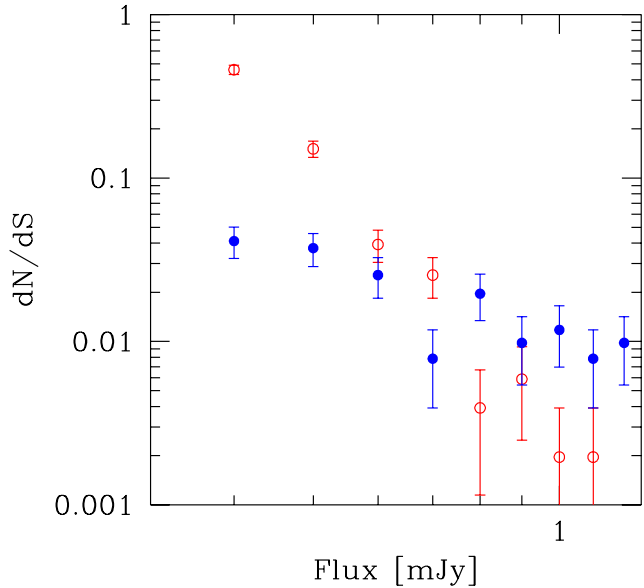}
\caption{Comparison of the differential counts of
``starburst-dominated'' ($F_{8\mu\rm m}/F_{24\mu\rm m} <  0.1$;
open (red) circles) vs ``AGN-dominated'' ($F_{8\mu\rm
m}/F_{24\mu\rm m} > 0.1$; filled (blue) circles) galaxies in the
$F_{24\mu\rm m}\ge 0.35\,$ sample of sources fainter than
$R=25.5$. Counts refer to $\Delta F_{24\mu\rm m} = 0.1\,$~mJy bins
and are normalized to the total number of objects (510) in the
overlapping MIPS-IRAC region. \label{fig:sbvsAGNcounts}}
\end{center}
\end{figure}

Indeed, Fig.~\ref{fig:sbvsAGNcounts} shows that the shapes of the
differential $24\,\mu$m counts for the two starburst and AGN
sub-populations are very different. ``AGNs'' dominate above
$\simeq 0.8\,$mJy, consistent with the observational evidence that
IR spectra (obtained with the IRS: Infrared Spectrograph on
Spitzer) for optically faint but sufficiently mid-IR bright
sources ($F_{24\mu\rm m}\simgt 0.75\,$mJy)  predominantly present
the typical shape of obscured AGNs (Houck et al. 2005; Yan et al.
2005; Weedman et al. 2006). On the other hand, ``starbursts'' are
found to prevail at fainter fluxes and therefore constitute the
dominant class in the present MIPS-IRAC $F_{24\mu\rm m}\simgt
0.35\,$mJy sample.

\subsection{Definition of the Sample}
The arguments presented throughout this Section show that
optically faint, 24$\, \mu\rm m$-selected objects typically
identify star-forming galaxies in the redshift interval $1.6\simlt
z\simlt 2.7$.

There are 793 $R>25.5$ MIPS galaxies with 24$\, \mu$m fluxes
brighter than 0.35~mJy in the area of approximately 3.97 square
degrees ($257.25^\circ \simlt \hbox{RA}(2000) \simlt
261.75^\circ$; 58.6$^\circ$ $\simlt \hbox{Dec} \simlt$
60.35$^\circ$) covered by KPNO data (cutting out the irregular
regions close to the borders of the 24$\mu \rm m$ field). This
region encloses the 2.85 square degrees where also 8$\, \mu$m data
is available. The sources selected in the above fashion correspond
to 7.4\% of the total number of objects (10,693) brighter than
$F_{24\mu\rm m}=0.35\,$mJy found in the same area and constitute
the sample which will be used in the following statistical
analyses.

It is worth noting that, while the adopted 24~$\mu$m limit ensures
completeness for what concerns the mid-IR selection of the sample,
the optical $R>25.5$ cut is somewhat arbitrary. In fact, the
studies presented in \S\,3.3 find sources in the $1.6\simlt
z\simlt 2.7$ redshift range having magnitudes brighter than our
chosen value. On the other hand, some of the sources with $R\sim
24$ observed by the above authors turned out to have lower
redshifts. One therefore has that a cut at $R=25.5$ ensures that
the overwhelming majority (if not all) of the selected objects lie
in the $z\sim 2$ range, while relaxing the optical magnitude limit
may lead to a contamination of the sample, while only adding a
modest fraction of sources (see also \S$\,$\ref{sec:clust} below).
For example, if we lower the limit to $R = 24.5$ we only add 139
objects to the adopted sample, corresponding to a fractional
increase of 17.5\%.

While such an incompleteness in the optical selection of the MIPS
sample is not expected to affect the clustering estimates as long
as the redshift distribution of $R>25.5$ sources does not greatly
differ in shape from that of {\it all} galaxies belonging to the
same population and endowed with mid-IR fluxes $F_{24\mu\rm m}\ge
0.35\,$mJy, the same does not hold when considering the space
density of such sources; in this case, the quantity quoted in
\S\,4 will have to be considered as a mere lower limit.

\section{Clustering Properties}\label{sec:clust}

\begin{figure}
\vspace{8cm}  
\includegraphics{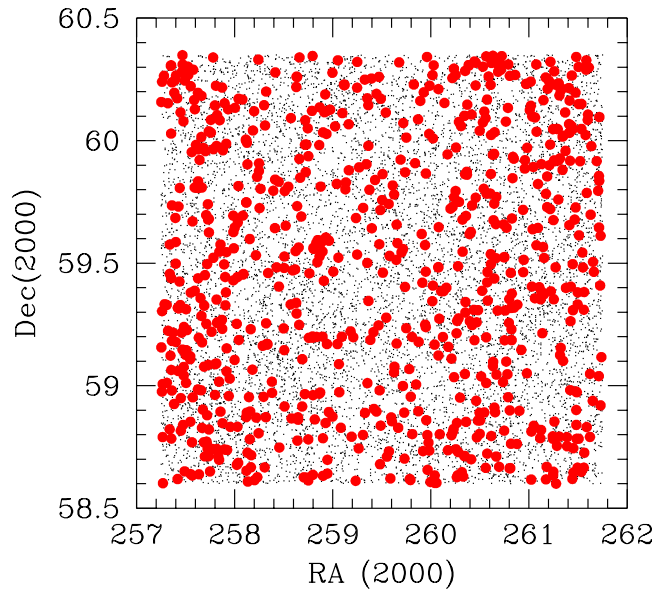} \caption{Sky distribution of the 793 optically
faint ($R>25.5$) sources in the $F_{24\mu\rm m}\ge 0.35$~mJy MIPS
sample found within the area of 3.97 square degrees covered by
KPNO data (red filled circles; see text for details). The
distribution of all the 24$\mu$m sources brighter than the same
flux limit and enclosed in the same region of the sky is also
shown for comparison (small black dots). \label{fig:radec}}
\end{figure}

\begin{figure}
\vspace{8cm}  
\includegraphics{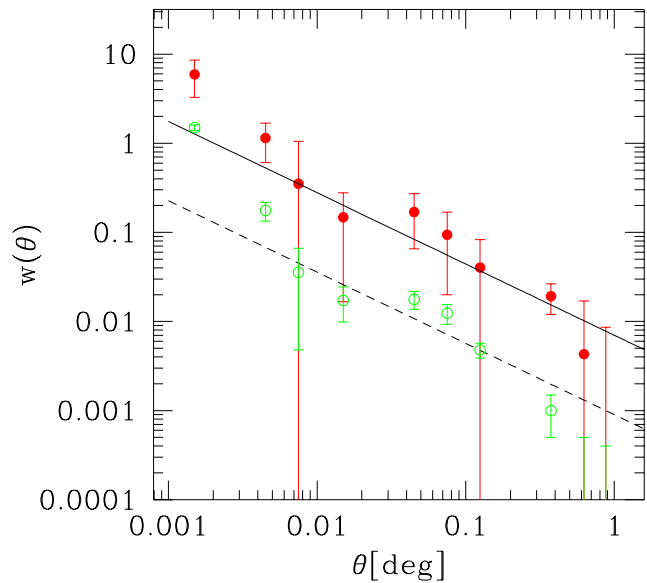} \caption{Angular
correlation function $w(\theta)$ for $R> 25.5$ sources (filled
circles) and for the whole $F_{24\mu \rm m}\ge 0.35$~mJy MIPS
sample (open circles). The solid and the dashed lines show the
best power-law fits to the data.
\label{fig:w_theta}}
\end{figure}

A mere visual inspection of the sky distribution of the $24\,\mu$m
sources with $R > 25.5$ which identify the sample presented in
\S\,3.4 (filled circles in Fig.~\ref{fig:radec}), indicates that
these objects are strongly clustered, much more than the sources
in the full $F_{24\mu \rm m}\ge 0.35$~mJy First Look Survey (small
dots).

The standard way to quantify the clustering properties of a
particular class of sources of unknown distance is by means of the
angular two-point correlation function $w(\theta)$ which measures
the excess probability of finding a pair in the two solid angle
elements $d\Omega_1$ and $d\Omega_2$ separated by an angle
$\theta$. In practice, $w(\theta)$ is obtained by comparing the
actual source distribution with a catalogue of randomly
distributed objects subject to the same mask constraints as the
real data.  We chose to use the estimator (Hamilton 1993)
\begin{eqnarray}
w(\theta) = 4\times \frac{DD\cdot RR}{(DR)^2} -1, \label{eq:xiest}
\end{eqnarray}
in the range of scales $10^{-3}\simlt\theta\simlt 1$ degrees. $DD$, $RR$ and
$DR$ are the number of data-data, random-random and data-random
pairs separated by a distance $\theta$. The random catalogue was
generated with twenty times as many objects as the real data set,
and the angular distribution of its sources was modulated
according to the MIPS coverage map, so that the instrumental
window function did not affect the measured clustering.

The resulting angular correlation function $w(\theta)$ for the
obscured ($R > 25.5$), $F_{24\mu \rm m}\ge 0.35$~mJy sources is
shown by the filled circles in Fig.~\ref{fig:w_theta}. %
The errors have been computed as:
\begin{eqnarray}
\delta w(\theta) = {1+ w(\theta) \over \sqrt{DD}}.
\label{eq:errw}
\end{eqnarray}
%

Since the distributions are clustered, this (Poisson) estimate for
the errors only provides a lower limit to the real uncertainties.
However, it can be shown  that over the considered range of angular
scales this estimate is close to those obtained from bootstrap
resampling (e.g. Willumsen, Freudling \& Da Costa, 1997).
On the other hand, the above estimate does not take into account
the uncertainties on the sample selection, which we are unable to
quantify but may be substantial. We tentatively allow for them by
doubling the Poisson errors in the following analysis and in the
Figures. {The above analysis was repeated by using the Landy \& Szalay
(1993) estimator, and we found virtually identical results.}

We have also investigated the possibility that the clustering
properties of high-$z$ starburst galaxies are contaminated by
obscured AGNs (see \S\,3.2), which may cluster differently.
However, removing candidate AGNs, i.e. objects with
$F_{8\mu\rm m}/F_{24\mu\rm m} >
0.1$ (or alternatively $F_{24\mu\rm m}\simgt 0.8\,$mJy)
{which make up for $\sim$20 per cent of the total sample},
leaves the angular correlation function
essentially unaffected, indicating that AGN-powered sources have
clustering properties similar to those of starbursts.

On the other hand, if we somewhat relax the optical magnitude
limit, e.g. we decrease it from $R> 25.5$ to $R> 24.5$, the
estimated $w(\theta)$ becomes significantly noisier, even though
the fraction of added sources is only 17.5\% of the original
sample (similar to that of ``AGNs''), suggesting that a
substantial portion of optically brighter, $F_{24\mu\rm m} \ge
0.35$ sources are at lower redshifts. Including still optically
brighter sources, the angular correlation function rapidly
decreases, approaching that obtained for the whole
$24\,\mu$m-selected sample (open circles in
Fig.~\ref{fig:w_theta}; in this case the associated errors simply
correspond to 1$\sigma$).

If we adopt the usual power-law form,
\begin{eqnarray}
w(\theta)=A\theta^{(1-\gamma)}, \label{eq:powerw}
\end{eqnarray}
the parameters $A$ and $\gamma$ can be estimated via a
least-squares fit to the data. However, given the large errors on
$w$, we choose to fix $\gamma$ to the standard value $\gamma=1.8$.
We then obtain $A=(7\pm 2)\times 10^{-3}$ (solid line in
Fig.~\ref{fig:w_theta}); the point on the top left-hand corner has
not been considered in our analysis because it corresponds to an
angular scale close to the resolution of the instrument and
therefore, despite the accuracy of the deblending technique
applied to produce the original MIPS catalogue, it may be affected
by source confusion.

The amplitude $A$ is about three times higher than that derived by
Fang et al. (2004) for a sample of IRAC galaxies selected at
8$\mu$m ($A\sim 2.34 \cdot 10^{-3}$), and about eight times higher
than that obtained for the whole $F_{24\mu \rm m} \ge 0.35$~mJy
MIPS dataset ($A=(9\pm 2)\cdot 10^{-4}$, dashed line in
Fig.~\ref{fig:w_theta}).

The angular correlation function is related to the spatial one,
$\xi(r,z)$, by the relativistic Limber equation (Peebles, 1980):
\begin{eqnarray}
w(\theta)=2\:\frac{\int_0^{\infty}\int_0^{\infty}F^{-2}(x)x^4\Phi^2(x)
\xi(r,z)dx\:du}{\left[\int_0^{\infty}F^{-1}(x)x^2\Phi(x)dx\right]^2},
\label{eqn:limber}
\end {eqnarray}
where $x$ is the comoving coordinate, $F(x)$ gives the correction
for curvature, and the selection function $\Phi(x)$ satisfies the
relation
\begin{eqnarray}
{\cal N}=\int_0^{\infty}\Phi(x) F^{-1}(x)x^2 dx=\frac{1}{\Omega_s}
\int_0^{\infty}N(z)dz, \label{eqn:Ndense}
\end{eqnarray}
in which $\cal N$ is the mean surface density, $\Omega_s$ is the
solid angle covered by the survey, and $N(z)$ is the number of
sources within the shell ($z,z+dz$).

If we make the simple assumption, consistent with the photometric
and spectroscopic information summarized in \S$\,$\ref{photspec},
that $N(z)$ is constant in the range $1.6 \simlt z \simlt 2.7$ and
adopt a spatial correlation function of the form
$\xi(r,z)=(r/r_0)^{-1.8}$, independent of redshift (in comoving
coordinates) in the considered interval, we obtain, for the
adopted cosmology, $r_0(z=1.6-2.7)=14.0_{-2.4}^{+2.1}$~Mpc. The
clustering radius increases (decreases) if we broaden (narrow) the
redshift range. If we instead adopt the redshift distribution
predicted by the model of Granato et al. (2004; see \S\,6) we get:
$r_0(z=1.6-2.7)=15.2_{-2.6}^{+2.3}$~Mpc. Note that the assumption
of a redshift independent comoving clustering radius is borne out
by observational estimates for optical quasars (Porciani et al.
2004; Croom et al. 2005) which -- according to the Granato et al.
(2004) model -- correspond to a later evolutionary phase of the
AGNs hosted by $24\,\mu$m sources.

The above value of $r_0$ is in good agreement with the estimates
obtained in the case of ultra-luminous infrared galaxies over $1.5
< z <3$ (Farrah et al. 2006a,b), and also matches that found by
Magliocchetti \& Maddox (1999) in the analysis of the clustering
properties of galaxies in the Hubble Deep Field North selected in
the same redshift range. Massive star-forming galaxies at $z\sim
2$ thus appear  to be amongst the most strongly clustered sources
in the Universe. Locally, their clustering properties find a
counterpart in those exhibited by radio sources (see e.g.
Magliocchetti et al. 2004) and are only second to those of rich
clusters of galaxies (e.g. Guzzo et al. 2000). The implications of
this result will be investigated in the next Sections.

Under the assumption of a uniform redshift distribution in the
range $1.6 \simlt z \simlt 2.7$  and for the adopted cosmology, the mean
comoving space density of sources with $R>25.5$ and $F_{24\mu\rm
m} \ge 0.35\,$mJy is:
%
\begin{equation}
\bar{n}_{\rm obs}(1.6 < z < 2.7) \sim 1.5\cdot 10^{-5}\ {\rm
Mpc}^{-3}. \label{nobs}
\end{equation}
%
%

\section{The Halo Occupation Number (HON)}

A closer look to Fig.~\ref{fig:w_theta} shows that a simple
power-law provides a good fit for the measured $w(\theta)$ only
over the range $0.007^\circ\simlt \theta\simlt 0.5^\circ$. {
Even though masked by large error bars, a hint of a steepening can
in fact be discerned} on the smallest angular scales and may also
be present at the largest angles probed by our analysis. The
small-scale steepening is intimately related to the way the
sources under exam occupy their dark matter haloes, issue which
will be dealt with throughout this Section via the so-called Halo
Occupation Scenario, while the steepening on large angular scales
is most likely due to the high redshift of these sources
($0.5^\circ$ in the adopted cosmology and for a redshift $z=2$
correspond to a scale of $\sim 45$~Mpc above which the real-space
correlation function rapidly approaches zero).

\subsection{Setting up the formalism}
The halo occupation function is defined as the probability
distribution of the number of galaxies brighter than some
luminosity threshold hosted by a virialized halo of given mass.
Within this framework, it is possible to show (see e.g. Peacock \&
Smith 2000 and Scoccimarro et al. 2001) that the distribution of
galaxies within their dark matter haloes determines the
galaxy-galaxy clustering on small scales. Since the distribution
of sources within their haloes in general depends on the
efficiency of galaxy formation, clustering measurements can
provide important insights on the physics of those objects
producing the signal. This approach has been successfully applied
in the past few years to a number of cases, from local galaxies
(Magliocchetti \& Porciani 2003; Zehavi et al. 2004) to higher
redshift sources such as COMBO 17 and Lyman Break galaxies (e.g.
Phleps et al. 2005; Hamana et al. 2003; Ouchi et al. 2005) and
quasars (Porciani, Magliocchetti \& Norberg 2004).

Our analysis follows the approach adopted by Magliocchetti \&
Porciani (2003), which is in turn based on the work by Scherrer \&
Bertschinger (1991) and Scoccimarro et al. (2001). The basic
quantity here is the halo occupation distribution function
$P_N(M)$ which gives the probability of finding $N$ galaxies
within a single halo as a function of the halo mass $M$. Given the
halo mass function $n(M)$ (number density of dark matter haloes
per unit comoving volume and $\log_{10}(M)$), the mean value of
the halo occupation distribution $N(M)\equiv \langle N \rangle(M)=
\sum_N N\, P_N(M)$ (which, from now on, we will call the halo
occupation number) completely determines the mean comoving number
density of galaxies in the desired redshift range:
\begin{equation}
n_{\rm gal}=\int n(M)\,N(M)\,dM\;. \label{eq:avern}
\end{equation}
Relations, analogous to eq.~(\ref{eq:avern}) and involving
higher-order moments of $P_N(M)$, can be used to derive the
clustering properties in the framework of the halo model. For
instance, the 2-point correlation function can be written as the
sum of two terms
\begin{equation}
\label{eq:xi} \xi(r)=\xi^{1h}(r)+\xi^{2h}(r)\;.
\end{equation}
The function $\xi^{1h}$, which accounts for pairs of galaxies
residing within the same halo, depends on the second factorial
moment of the halo occupation distribution $\sigma(M)=\langle
N(N-1) \rangle(M)$ and on the spatial distribution of galaxies
within their host haloes $\rho({\bf x}|M)$, normalized in such a
way that $\int_0^{r^{\rm vir}} \rho({\bf y}|M)d^3y=1$ where
$r^{\rm vir}$ is the virial radius which is assumed to mark the
outer boundary of the halo:
\begin{eqnarray}
\xi^{1h}(r)=\!\!\int \frac{n(M)\,\sigma(M)}{n_{\rm gal}^2}\,dM
\int \rho({\bf x}|M)\,\rho({\bf x}+{\bf r}|M)\,d^3x.
\label{eq:xi1h}
\end{eqnarray}
On the other hand, the term $\xi^{2h}$, which takes into account
the contribution to the correlation function coming from galaxies
in different haloes, depends on both $N(M)$ and $\rho({\bf x}|M)$ as
follows
\begin{eqnarray}
\xi^{2h}(r)\!\!\!&=&\!\!\!\!\! \int \frac{n(M_1)\,N(M_1)}{n_{\rm
gal}}
\,dM_1 \int \frac{n(M_2)\,N(M_2)}{{n_{\rm gal}}}\,dM_2 \nonumber \\
&\times&\!\!\!\!\! \int\!\!  \rho({\bf x}_1|M_1)\, \rho({\bf
x}_2|M_2)\, \xi({\bf r}_{12}| M_1, M_2) \,d^3r_1\,d^3 r_2 ,
\label{eq:xi2h}
\end{eqnarray}
where $\xi(r| M_1, M_2)$ is the cross-correlation function of
haloes of mass $M_1$ and $M_2$, ${\bf x}_i$ denotes the distance
from the centre of each halo, and ${\bf r}_{12}$ is the separation
between the haloes.

For separations smaller than the virial radius of the typical
galaxy host halo, the 1-halo term dominates the correlation
function, while the 2-halo contribution is the most important one
on larger scales. In this latter regime $\xi(r| M_1, M_2)$ is
proportional to the mass autocorrelation function, i.e. $\xi(r|
M_1, M_2)\simeq b(M_1)\,b(M_2)\,\xi_{dm}(r)$, where $b(M)$ is the
linear bias factor of haloes of mass $M$. Note that all the
different quantities introduced in this Section depend on the
redshift $z$ even though we have not made it explicit in the
equations.

In order to use the halo model to study the galaxy clustering, one
has to specify a number of functions describing the statistical
properties of the population of dark matter haloes. In general,
these have either been obtained analytically and calibrated
against N-body simulations, or directly extracted from numerical
experiments.

For the mass function and the linear bias factor of dark matter
haloes we adopt here the model by Sheth \& Tormen (1999), while we
write the two-point correlation function of dark-matter haloes as
(see e.g. Porciani \& Giavalisco 2002; Magliocchetti \& Porciani
2003)
\begin{eqnarray}
\xi(r|M_1,M_2)\!\!=\!\!\left\{\begin{array}{ll}
\!\!\!\!\xi_{dm}(r)b_1(M_1)b_2(M_2) &\!\!\!\! {\rm if}\; r\ge r^{\rm vir}_1\!\!+r^{\rm vir}_2\\
\!\!\!\!-1 & {\rm otherwise},
\end{array}\right.
\label{eq:xihalo}
\end{eqnarray}
where the mass autocorrelation function, $\xi_{dm}(r)$, is
computed with the method of Peacock \& Dodds (1996) and the above
expression takes into account the spatial exclusion between haloes
(i.e. two haloes cannot occupy the same volume). We will also
assume that the distribution of galaxies within their haloes
traces that of the dark matter and we adopt for $\rho({\bf x}|M)$
a Navarro, Frenk \& White (1997) profile with a concentration
parameter obtained from eqs.~(9) and (13) of Bullock et al.
(2001). In fact, Magliocchetti \& Porciani (2003) showed that NFW
profiles are well suited to describe the correlation function of
local 2dF galaxies. We note however that the uncertainties
associated to our estimates of $w(\theta)$ do not allow us to
discriminate between different forms for $\rho({\bf x}|M)$ as long
as they are sensible ones ({e.g. profiles of the form $\rho\propto r^{-\beta}$
with $2\simlt\beta\simlt 3$, $\beta=2$ corresponding to
the singular isothermal sphere case)}.

\begin{figure}
\vspace{8cm}  
\includegraphics{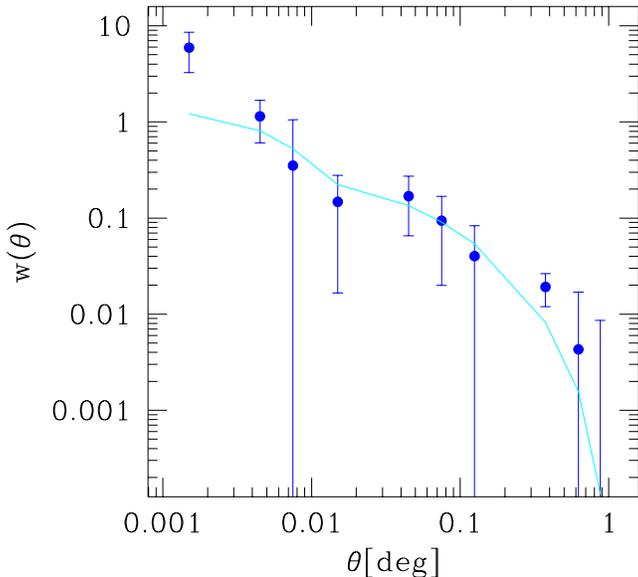} \caption{Best fit to the
observed angular correlation function $w(\theta)$ of our sources,
in the HON framework for $\alpha=0.2$, $M_{\rm min}=10^{13.4}
M_\odot$, $\rm log_{10}(N_0)=-0.3$. The point on the smallest
angular scale has not been taken into account, because of
uncertainties in source deblending. \label{fig:whon}}
\end{figure}

The final key ingredient needed to describe the clustering
properties of a class of galaxies is their halo occupation
function $P_N(M)$. In the ideal case $P_N(M)$ is entirely
specified by the knowledge of all its moments which in principle
can be observationally determined by studying galaxy clustering at
any order. Unfortunately this is not feasible in practice, as
measures of the higher moments of the galaxy distribution get
extremely noisy for $n>4$ even for local 2-dimensional catalogues
(see e.g. Gazta\~{n}aga, 1995 for an analysis of the APM survey).
On the other hand, the present work relies on measurements of the
two-point correlation function, which only depends on the first
two moments of the halo occupation function, $N(M)$ and
$\sigma(M)$.

Following Porciani et al. (2004; see also
Magliocchetti \& Porciani 2003 and Hatton et al. 2003) we
parameterize these quantities as:
\begin{eqnarray}
N(M) = \left\{\begin{array}{ll} N_0(M/M_{\rm min})^{\alpha} &
\rm{if}\ M \ge  M_{\rm min} \\
0 & \rm {if} \ M<M_{\rm min}
\end{array}\right.
\label{eq:Ngal}
\end{eqnarray}
and
\begin{eqnarray}
\sigma(M)=\beta(M)^2 N(M),
\end{eqnarray}
where $\beta(M)=0,{\rm log}(M/M_{\rm min})/{\rm log}(M/M_0),1$,
respectively for $N(M)=0$, $N(M)<1$ and $N(M)\ge 1$. The
operational definition of $M_0$ is such that $N(M_0)=1$ (see e.g.
Porciani et al. 2004), while $M_{\rm min}$ is the minimum mass of
a halo able to host a source of the kind under consideration. More
and more massive haloes are expected to host more and more
galaxies, justifying the assumption of a power-law shape for the
halo occupation number. As already pointed out in Porciani et al.
(2004), eq.~(\ref{eq:Ngal}) is more general than the commonly used
$N(M)=(M/M_0)^\alpha$ which, for $\alpha=0$, automatically implies
$N(M)=1$ at any $M>M_0$.

As for the variance $\sigma(M)$, we note that the high-mass value
for $\beta(M)$ simply reflects the Poisson statistics, while the
functional form at intermediate masses (chosen to fit the results
from semi-analytical models -- see e.g. Sheth \& Diaferio 2001;
Berlind \& Weinberg 2002 -- and hydrodynamical simulations --
Berlind et al. 2003) describes the (strongly) sub-Poissonian
regime. We assume the various quantities describing the HON not to
vary in the considered redshift range. Although a simplification,
this choice is partially justified by the results obtained for
other extragalactic sources sampling the same redshift range of
our dataset (e.g. quasars, see Porciani et al. 2004), which indeed
show the relevant parameters associated to $N(M)$ to stay constant
with look-back time.

\begin{figure}
\vspace{8cm}  
\includegraphics{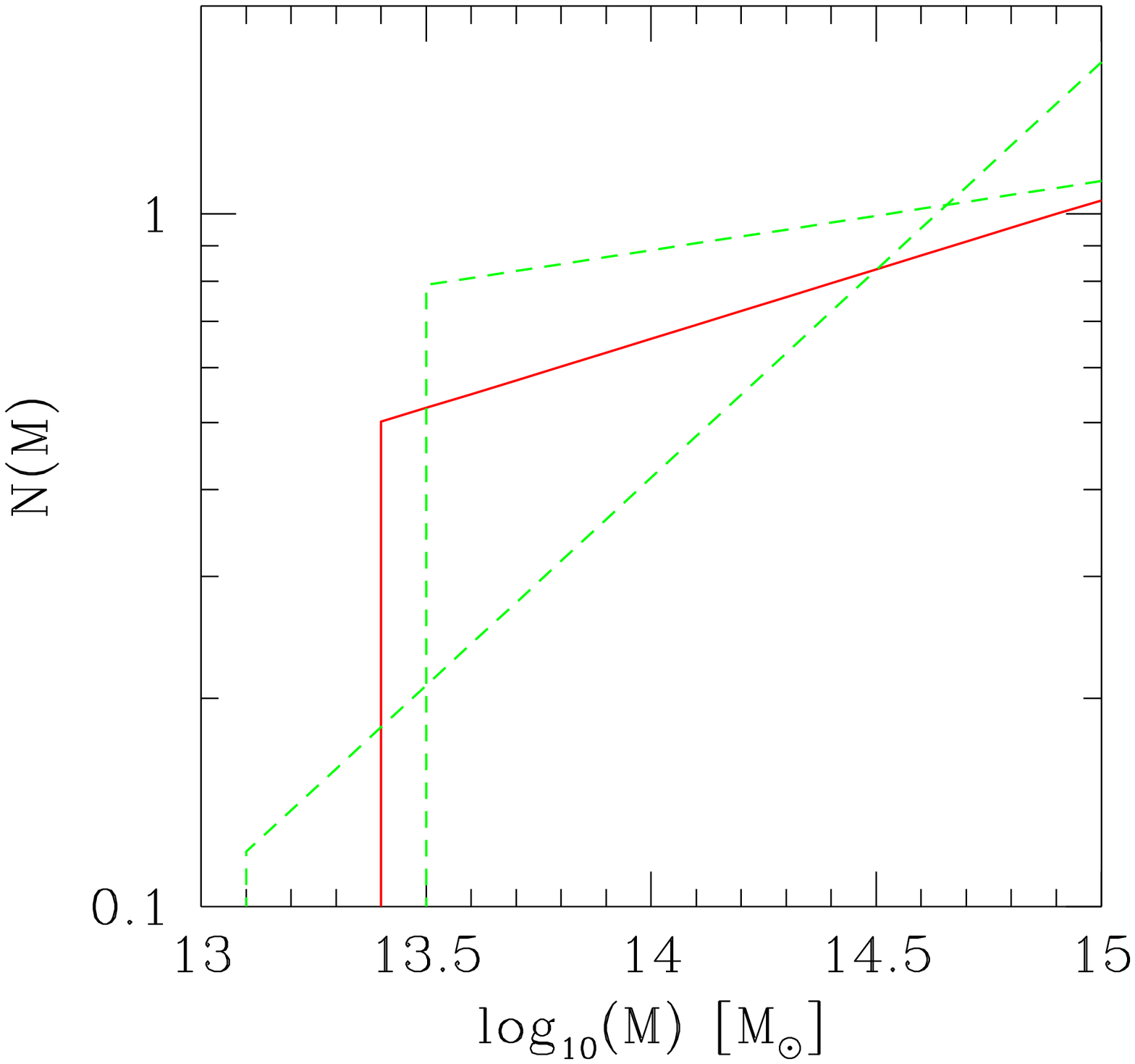}
\caption{Average number of MIR-bright ($F_{24\mu\rm m}\ge
0.35$~mJy), optically faint ($R>25.5$), galaxies per dark matter halo of specified
mass $M$ in the redshift range $z\simeq [1.6-2.7]$. The solid line
corresponds to the best-fit values of
eq.~(15), while the dashed lines
correspond to cases with $\Delta\chi^2=1$ (see text).  
\label{fig:HON}}
\end{figure}

\subsection{Results}
In the application of the HON formalism to the present sample we
allowed the parameters of eq.~(\ref{eq:Ngal}) to vary within the
following ranges:
\begin{eqnarray}
& 0 \le \alpha\le 2  \nonumber\\
& 10^{11}\; M_\odot\le M_{{\rm min}} \le 10^{14}\; M_\odot \\
& -2\le \log_{10}(N_0)\le 1\ .\nonumber
\end{eqnarray}
Values for these parameters have been determined through a
minimum $\chi^2$ technique by fitting the observed $w(\theta)$
(except for the smallest angular scale point, cfr.
\S$\,$\ref{sec:clust}).
The angular correlation function was computed from
eq.~(\ref{eqn:limber}), with $\xi(r)$ given by eq.~(\ref{eq:xi})
and the redshift distribution yielded by the Granato et al. (2004)
model (see \S$\,$\ref{sec:model}).

We find that the best fit to the $w(\theta)$ alone is obtained for:
\begin{eqnarray}
\alpha=0.2^{+0.7}_{-0.2}\;\;\;\;\;\;\;\;\;\;\;\;\;\;\;\;\;\;\;\;\;\;\;\;\;
\nonumber\\
{\log}_{10}(M_{\rm min}/M_\odot)=13.5^{+0.2}_{-0.4}\;\;, \\
{\log}_{10}(N_0)=-0.3^{+0.2}_{-1.7}\;\;\;\;\;\;\;\;\;\;\nonumber
\end{eqnarray}
where the quoted errors correspond to $\Delta\chi^2=1$. We note however
that the sampled correlation function bins are not completely independent,
so that the error estimate is only indicative.

The parameters are correlated with each other. In particular,
higher values for $M_{\rm min}$ correspond to lower values of
$\alpha$ (see Fig.~\ref{fig:HON}). Furthermore, as $N_0$ is always
found to be less than 1 and the index $\alpha$ is rather flat, on
average there is less than one of such star-forming galaxies per
dark matter halo (sub-Poissonian regime) except at the
highest/cluster-like mass scales.

It may be noted that the data on $w(\theta)$ can only effectively constrain
$M_{\rm min}$, while allowing for relatively broad ranges in the case of
$N_0$ and $\alpha$. This can be easily understood since, for
typical halo masses $>>M^*$ the halo bias function $b(M)$ is a
steep function of $M$, so that even small variations of $M_{\rm
min}$ result in large variations of the predicted $w(\theta)$ on
intermediate-to-large angular scales. On the other hand, $N_0$
can only be constrained by data on small angular scales
(one-halo regime) via the theoretical variance $\sigma(M)$. But
the one-halo regime is represented by just one data point and
furthermore, as long as the regime is sub-Poissonian, the predicted
one-halo correlation function is only mildly dependent on this
quantity.

Additional constraints on the three parameters characterizing the
HON [eq.~(\ref{eq:Ngal})] can be obtained from the estimated
comoving number density of our sources. In fact, any HON model
must simultaneously be able to reproduce both the first and second
moment of the galaxy distribution, i.e. both the clustering
properties and the observed number density of sources in a
specified sample. As discussed in \S\,4, the estimate in eq.~(6)
is expected to provide a lower limit to the number density of
high-redshift star-forming galaxies with $F_{24\mu\rm m}\ge 0.35$,
although with a substantial uncertainty, mostly related to our
poor knowledge of the redshift distribution. If we then require
$n_{\rm gal}$ in eq. (\ref{eq:avern}) to be $\ge 6\cdot 10^{-6}\ {\rm Mpc}^{-3}$
(i.e. allow for an uncertainty of a factor of 2.5)
the permitted ranges for the parameters narrow down becoming:
\begin{eqnarray}
\alpha=0.2^{+0.4}_{-0.1}\;\;\;\;\;\;\;\;\;\;\;\;\;\;\;\;\;\;\;\;\;\;\;\;\;
\nonumber\\
{\log}_{10}(M_{\rm min}/M_\odot)=13.4^{+0.1}_{-0.3} \\
{\log}_{10}(N_0)=-0.3^{+0.2}_{-0.6}\ ,
\;\;\;\;\;\;\;\;\;\;\nonumber \label{eq:bestvalues}
\end{eqnarray}
which are the best-fit values for the HON [eq.~(\ref{eq:Ngal})]
satisfying both the clustering and the number density requirement.
We note that, while as expected the range for $M_{\rm min}$ is
basically unaffected, the constraint on $n_{\rm gal}$ greatly
shrinks the allowed region for $N_0$ by cutting all those values
which would have produced too few sources.
The above best-fit values do not change significantly if instead
of the $N(z)$ deriving from the Granato et al. (2004) model we use
the flat redshift distribution introduced in \S\,3; in this case
we get $\alpha=0.3^{+0.3}_{-0.2}$; ${\log}_{10}(M_{\rm
min}/M_\odot)=13.3^{+0.1}_{-0.2}$;
${\log}_{10}(N_0)=-0.4^{+0.3}_{-0.3}$.

The theoretical angular correlation function corresponding to the
above best-fit HON parameters is compared to the data in
Fig.~\ref{fig:whon}. The model correctly describes both the
overall amplitude of $w(\theta$) and the rise on angular scales
$\simlt 10^{-2}$~deg determined by the one-halo regime. The
corresponding Halo Occupation Number of high-redshift star-forming
galaxies is presented in Fig.~\ref{fig:HON}, {which shows that the sources under
exam are always associated to very massive structures, identifiable with
groups-to-clusters of galaxies. As it is also possible to notice, such galaxies are reasonably
common in those massive structures, with an average of $\sim 0.5-1$ object per group, where the
upper value is found in correspondence of the highest masses probed by our analysis.
The implications of these results will be investigated in the next Section, when discussing the
nature of optically faint objects as selected at 24$\mu$m.}


%
\begin{figure}
\vspace{8cm}  
\includegraphics{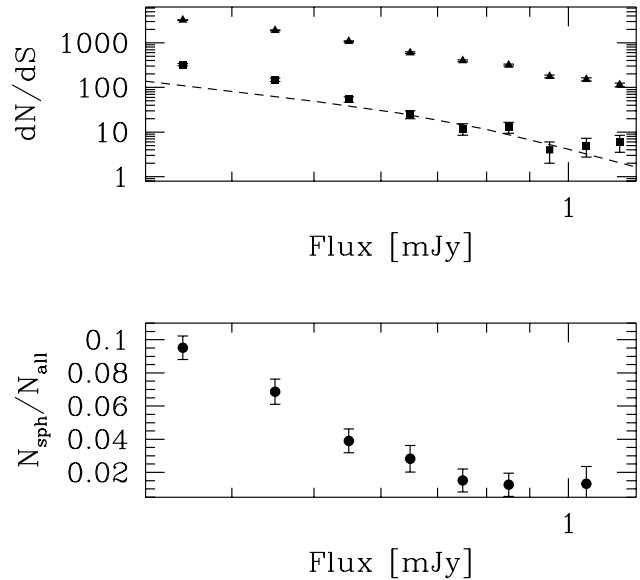}
\caption{{Number of $24\,\mu$m sources fainter than $R=25.5$
(filled squares) and of all 24$\mu$m MIPS sources counted, in
$\Delta F_{24\mu\rm m}=0.1\,$mJy bins, in the $3.97\,\hbox{deg}^2$
area covered by KPNO data.}  
The lower panel
represents the ratio between the above quantities as a function of
the 24$\mu$m flux, while the dashed line in the top panel shows
the predictions by Silva et al. (2004; 2005) for high-$z$
proto-spheroidal galaxies. \label{fig:N_S}}
\end{figure}
\begin{figure}
\vspace{8cm}  
\includegraphics{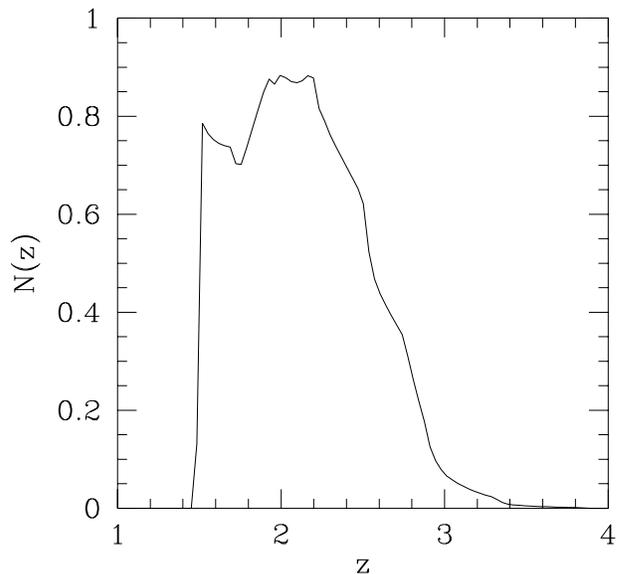}
\caption{Redshift distribution of dusty proto-spheroidals with
fluxes $F_{24\mu\rm{m}}\ge 0.35\,$mJy, normalized to unity, 
obtained following Silva et al. (2004; 2005). \label{fig:N_z}}
\end{figure}

\section{Nature of the sources}\label{sec:model}

Before the {\it Spitzer} survey data became available, Silva et
al. (2004; 2005) worked out detailed predictions for the counts
and the redshift distributions of IR sources. In particular they
predicted that {for $F_{24\rm \mu m}\simgt 0.35$~mJy}, MIPS
surveys would have comprised a small, but significant (8--10\%)
fraction of objects in the redshift range $\sim 1.5--2.6$ (with a tail
extending up to $z\simgt 3$; cfr. their Figure~27). {At the
flux limit ($F_{24\rm \mu m}\ge 83\,\mu$Jy) of the MIPS survey of
the Chandra Deep Field South (Papovich et al. 1984), Silva et al.
(2004; 2005) predicted a surface density of proto-spheroids at $z
\ge 1.5$ of $\simeq 1\,\hbox{arcmin}^{-2}$, i.e. amounting to $\simeq
22\%$ of the observed total surface density.} This prediction was
at variance with those of reference phenomenological models which,
for $F_{24\mu\rm m} \ge 83\,\mu$Jy, yielded (see Figure~2 of
P\'erez-Gonz\'alez et al. 2005 {and Figure~6 of Caputi et al.
2006}) either very few (Chary et al. 2004) or almost 50\% (Lagache
et al. 2004) sources at $z\simgt 1.5$. The redshift distribution
by P\'erez-Gonz\'alez et al. (2005) for $F_{24\mu\rm m} \ge
83\,\mu$Jy, based primarily on photometric redshifts for starburst
templates, has about 24\% of sources at $z\ge 1.5$. {This
result was recently confirmed, within the errors, by the work of
Caputi et al. (2006) who found $\simeq 28\%$ of sources to lie in
that $z$ range.}


{The basic difference between the work of Silva et al.
(2004; 2005) and that of the other quoted models is that, while
Chary et al. (2004) and Lagache et al. (2004) adopt a purely
empirical/phenomenological approach to describe the high redshift population
of galaxies selected at 24$\mu$m (by e.g. evolving the local luminosity function
in luminosity and/or density), Silva et al. (2004; 2005) consider a more
physically grounded picture.}
In fact, according to Silva et al. (2004; 2005) the $z
\simgt 1.5$ population corresponds to massive proto-spheroidal
galaxies in the process of forming most of their stars in a
gigantic starburst, whose evolution is described by the physical
model of Granato et al. (2004). This population is not represented
in the local far-IR luminosity function, since local massive
spheroids are essentially dust-free. We refer the interested
reader to the Granato et al. (2004) paper for a full account of
the physical justification and a detailed description of the
model. Here we provide a short summary of its main features,
focusing on the aspects which are relevant to the present
discussion.

\subsection{Overview of the Granato et al. (2004) model}

The model adopts the standard hierarchical clustering framework
for the formation of dark matter haloes. It focuses on the
redshift range $z \ga 1.5$, where a good approximation of the halo
formation rates is provided by the positive term in the cosmic
time derivative of the cosmological mass function (e.g., Haehnelt
\& Rees 1993; Sasaki 1994). The simulations by Wechsler et al.
(2002), and Zhao et al. (2003a,b) show that the growth of a halo
occurs in two different phases: a first regime of fast accretion
in which the potential well is built up by the sudden mergers of
many clumps of comparable mass; and a second regime of slow
accretion in which mass is added in the outskirts of the halo,
without affecting the central region where the galactic structure
resides. This means that, even at high redshift, once created the haloes harboring a
massive elliptical galaxy are rarely destroyed and
get incorporated within groups and clusters of galaxies.

The physics governing the evolution of the baryons is much more
complex. The main features of the model can be summarized as
follows (see Granato et al. 2004, Cirasuolo et al. 2005, Lapi et
al. 2006). During or soon after the formation of the host dark
matter halo, the baryons falling into the newly created potential
well are shock-heated to the virial temperature. The hot gas is
(moderately) clumpy and cools quickly especially in the denser
central regions, triggering a strong burst of star formation. The
radiation drag due to starlight acts on the gas clouds, reducing
their angular momentum. As a consequence, a fraction of the cool
gas can fall into a reservoir around the central super-massive
black hole, and eventually accretes onto it by viscous
dissipation, powering the nuclear activity. The energy fed back to
the gas by supernova explosions and black hole activity regulates
the ongoing star formation and the black hole growth.

Initially, the cooling is rapid and the star formation is very
high; thus the radiation drag is efficient in accumulating mass
into the reservoir. The black hole starts growing from an initial
seed with mass $\sim 10^2\, M_{\odot}$ already in place at the
galactic center. Since there is plenty of material in this phase,
the accretion is Eddington (or moderately super-Eddington) limited
(e.g., Small \& Blandford 1992; Blandford 2004). This regime goes
on until the energy feedback from the black hole is strong enough
to unbind the gas from the potential well, a condition occurring
around the peak of the accretion curve. Subsequently, the star
formation rate drops substantially, the radiation drag becomes
inefficient, the storage of matter in the reservoir and the
accretion onto the black hole decrease by a large factor. The drop
is very pronounced for massive haloes, $M_{\mathrm{vir}}\ga
10^{12}\, M_{\odot}$, while for smaller masses a smoother
declining phase can continue for several Gyrs, and the black hole
and stellar masses can further increase by a substantial factor.

Before the peak, radiation is highly obscured by the surrounding
dust. In fact, these proto-galaxies are extremely faint in the
rest frame UV/optical/near-IR and are more easily selected at
far-IR to mm wavelengths. Nuclear emission is heavily obscured
too, but since absorption significantly decreases with increasing
X-ray energy of the photons, this may be detected in hard X-ray
bands (Alexander et al. 2005; Borys et al. 2005; Granato et al.
2006). The mid-IR region may also be particularly well suited to
detect such an emission, because the dust temperature in the
nuclear torus, hotter than that of the interstellar medium, makes
this component more prominent, and the optical depth is relatively
low.

\subsection{Model versus observations}

The 24$\mu$m counts of candidate high-$z$ proto-spheroidal
galaxies predicted by Silva et al. (2004; 2005) are compared, in
Fig.~\ref{fig:N_S}, with those of MIPS sources fainter than
$R=25.5$. The total 24$\mu$m counts for the MIPS sample are also
shown for comparison.

The fraction of optically faint sources decreases from $\simeq
10\%$ at the lowest 24$\mu$m fluxes to less than 5\% at the
brightest ones. The decrement of this fraction with increasing flux
is slowed down or halted above $\simeq 0.8\,$mJy, when the `AGN'
contribution takes over.

The dashed line in Fig.~\ref{fig:N_S} represents the predictions
by Silva et al. (2004; 2005), based on the Granato et al. (2004)
model, for the counts at 24$\mu$m of dusty proto-spheroidals
undergoing intense star formation. It must be noted that Silva et
al. adopted a highly simplified description of the very complex
source spectra in the relevant rest-frame frequency range, where
strong emission and absorption features present a broad
distribution of equivalent widths. Also, although the model
explicitly predicts a significant nuclear activity with an
exponential growth of the central black hole mass during the
active star forming phase, nuclear emission was neglected in the
calculations by Silva et al. (2004; 2005). Thus, an accurate match
of the observed counts cannot be expected. Still, the observed
counts of optically faint sources are remarkably close to the
predictions, suggesting that objects in the present sample are
high-$z$ proto-spheroidal galaxies. The redshift distribution of
those sources brighter than 0.35 mJy at 24$\mu$m
(Fig.~\ref{fig:N_z}), computed by following Silva et al. (2004;
2005), matches the redshift range estimated in the previous
sections on the basis of photometric and spectroscopic evidences.

Following again Silva et al. (2004; 2005) we obtain that $\sim 5\%$ of
proto-spheroidal galaxies with $F_{24\mu{\rm m}} \ge 0.35\,$mJy
have $850\,\mu$m flux $\ge 10\,$mJy, consistent with the fact that
none of the four sources in our sample lying in the 151
arcmin$^{2}$ area surveyed with SCUBA by Sawicki \& Webb (2005)
to the above $850\,\mu$m flux limit was detected. The model
predicts most (80--90\%) of proto-spheroidal galaxies with
$F_{24\mu{\rm m}}\ge 0.35$~mJy to have $F_{850\mu{\rm m}}\ge
1$~mJy, but these are only a small fraction (2--4\%) of
${850\mu{\rm m}}$ sources at this flux limit.

We note that such a spread in the predicted $850\,\mu$m fluxes
reflects the spread in star formation rates (SFRs). The $\sim 5\%$
of proto-spheroidal galaxies selected at 24~$\mu$m with
$F_{850\mu{\rm m}} \ge 10\,$mJy are predicted to have $\hbox{SFR}
\simgt 1000\,M_\odot\,\hbox{yr}^{-1}$, while about 20--30\% of the
sample sources should have $\hbox{SFR} \simgt
500\,M_\odot\,\hbox{yr}^{-1}$, and about 90\% are characterized by
$\hbox{SFR}\simgt 100\,M_\odot\,\hbox{yr}^{-1}$. Most of the
sources have $\hbox{SFR} \sim \hbox{few}\times
100\,M_\odot\,\hbox{yr}^{-1}$.

The median halo mass, estimated from the model, is
$\log_{10}(M_{\rm vir}/M_\odot) \simeq 12.7$. The corresponding
peak SFR ranges from 550 to $\sim 800\, M_\odot\,\hbox{yr}^{-1}$
for virialization redshifts ranging from 3 to 4, i.e. is
substantially higher than the typical SFRs of the $24\,\mu$m
sources. This means that, according to the Granato et al. (2004)
model, the $24\,\mu$m selection preferentially identifies sources
in the phase when the effect of feedbacks has begun to damp the
SFR, at the same time decreasing the optical depth, which, at
earlier times, is very high even at rest-frame  mid-IR
wavelengths. In this phase the active nucleus is approaching its
maximum luminosity and can therefore show up at relatively bright
flux densities for a relatively short time, while the starburst
luminosity is far higher than that of the AGN over most of its
lifetime.

Also, the estimated halo mass is substantially lower than that
accounting for the clustering properties (cfr. \S\,5). According
to the model, this difference is to be expected since the
starburst phase of these haloes has a lifetime which is shorter
than the Hubble time by a factor of $\sim 5$ at $z\sim 2$. This
means that starburst galaxies work as beacons signalling the
presence of larger haloes, typically hosting $\sim 5$ galactic
haloes with $\log_{10}(M_{\rm vir}/M_\odot) \simeq 12.7$, out of
which only one is seen at $24\,\mu$m.

\section{Conclusions}

We have found that optically very faint ($R>25.5$) galaxies
selected at $24\,\mu$m ($F_{24\mu{\rm m}} \ge 0.35\,$mJy) are very
strongly clustered. Their two-point angular correlation function
has an amplitude which is about 8 times higher than that found for
the full $F_{24\mu{\rm m}} \ge 0.35\,$mJy sample, and about 3
times higher than the one estimated by Fang et al. (2004) for a
sample selected at $8\,\mu$m. Radio, sub-mm, mid-IR, and optical
photometric data converge in indicating that these sources are
very luminous star-forming galaxies set at redshifts $1.6 \simlt z
\simlt 3$. Spectroscopic redshifts for sources with similar
photometric properties fall in the range $1.6 \le z \le 2.7$. If
sources have a relatively flat distribution in the above redshift
interval, and we adopt the conventional power-law representation
for the spatial two-point correlation function,
$\xi(r,z)=(r/r_0)^{-1.8}$, we obtain, for the adopted cosmology, a
comoving clustering radius of
$r_0(z=1.6-2.7)=14.0_{-2.4}^{+2.1}\,$Mpc, implying that these
sources are amongst the most strongly clustered objects in the
universe. This result is in good agreement with the estimates for
ultra-luminous infrared galaxies over the redshift range $1.5 < z
<3$ obtained by Farrah et al. (2006a,b) by selecting sources with
$F_{8\mu{\rm m}} \ge 0.4\,$mJy, $R >22$ and bumps in either the
4.5 or the $5.8\,\mu$m IRAC channel: $r_0=14.4\pm 1.99
h^{-1}\,$Mpc for the $2<z<3$ sample (bump in the $5.8\,\mu$m
channel) and $r_0=9.40\pm 2.24 h^{-1}\,$Mpc for the $1.5<z<2.0$
sample (bump in the $4.5\,\mu$m channel).

The halo model provides a good fit of the observed angular
correlation function for a minimum halo mass of $\simeq
10^{13.4}\, M_\odot$. The number density of haloes above this mass
if fully consistent with that of our sources, therefore providing an
independent test of the results derived from the $w(\theta)$ alone.

At rest-frame wavelengths $\sim 8\mu$m, (corresponding to the
selection wavelength for the typical redshifts $z\sim 2$ of our
sources) both the direct starlight and the interstellar dust
emission are relatively low, so that nuclear activity can more
easily show up. In fact, indications that optically faint {\it
Spitzer} sources with $F_{24\mu{\rm m}} \simgt 1\,$mJy are AGN
dominated have been reported (Houck et al. 2005; Weedman et al.
2006). Comparing the redshift dependencies of the $8\mu{\rm
m}/24\mu{\rm m}$ flux ratios corresponding to the SEDs of well-known galaxies
such as Arp~220 (starburst galaxy) and of Mkn~231 (obscured AGN)
we find that, over the redshift range of interest
here, starburst galaxies are expected to have $F_{8\mu{\rm
m}}/F_{24\mu{\rm m}} < 0.1$ and AGN-dominated sources $F_{8\mu{\rm
m}}/F_{24\mu{\rm m}} > 0.1$. Adopting this criterion, we find that
in our sample the latter sources dominate for $F_{24\mu{\rm m}}
\ge 0.8\,$mJy, while starburst galaxies prevail at fainter fluxes
comprising $\simeq 80\%$ of the sample. No significant difference
in the clustering properties of the two sub-populations was
detected.

Our optically faint sources that, as argued above, are most likely
at $1.6 \le z \le 2.7$, comprise $\simeq 7.4\%$ of the complete
$F_{24\mu{\rm m}} \ge 0.35\,$mJy sample. This fraction is
remarkably close to the prediction by Silva et al. (2004; 2005).
These authors pointed out that the physically grounded
evolutionary model for massive spheroidal galaxies by Granato et
al. (2004) implies that the active star-forming phase of these
objects had to show up in $24\,\mu$m MIPS surveys and estimated
that they constitute $\simeq 8$--10\% of all sources brighter than
$F_{24\mu{\rm m}} \simeq 0.35\,$mJy and cover the redshift range
1.5--2.6 (with a tail extending up to $z\simgt 3$; cfr. Figure~27
of Silva et al. 2004). At the flux limit ($83\,\mu$Jy) of the
$24\,\mu$m MIPS survey of the Chandra Deep Field South, the model
predicts a surface density of $\simeq 1.0\,\hbox{arcmin}^{-2}$ for
sources at $z>1.5$ ($\simeq 22\%$ of the total surface density of
sources brighter than that flux limit), in nice agreement with the
observational determinations by P\'erez-Gonz\'alez et al. (2005)
and Caputi et al. (2006), yielding fractions of 24\% and 28\%
(corresponding to surface densities of $\simeq
1.1$--$1.3\,\hbox{arcmin}^{-2}$.


The photometric (radio, sub-mm, mid-IR, optical) properties of our
sources are all consistent with their interpretation in terms of
massive star-forming proto-spheroidal galaxies. The Granato et al.
(2004) model features a tight connection between star-formation
activity and growth of the active nucleus hosted by the galaxy. It
thus naturally accounts for the indications of a dominant AGN
contribution for $F_{24\mu{\rm m}} \ge 0.8\,$mJy.

According to this model, most sources have high, but not extreme,
star-formation rates (SFRs), $\hbox{SFR} \sim \hbox{few}\times
100\,M_\odot\,\hbox{yr}^{-1}$, well below the peak SFR reached
during their evolution. This is because during the most active
star-forming phases the optical depth of the interstellar medium
is very high even at rest-frame wavelengths of $\sim 8\mu$m. Thus,
the $24\,\mu$m selection preferentially singles out sources in the
phase when the effect of feedbacks has begun to damp the SFR, at
the same time substantially decreasing the optical depth. In this
phase, the active nucleus is approaching its maximum luminosity
and shows up, for a short time, at the brightest flux levels, as
is indeed observed.

The median galactic halo mass, estimated from the model, is
$\log_{10}(M_{\rm vir}/M_\odot) \simeq 12.7$, sensibly lower than
that accounting for the clustering properties. This difference is
due to the fact that the lifetime of the starburst phase of these
haloes is shorter (by a factor $\simeq 5$) than the Hubble time,
so that starburst galaxies work as beacons signalling the presence
of larger haloes, typically hosting $\simeq 5$ galactic haloes
with $\log_{10}(M_{\rm vir}/M_\odot) \simeq 12.7$, out of which
only one is bright enough at $24\,\mu$m to meet the observational
selection.

Once we account for this lifetime effect, the comoving density of
proto-spheroidal galaxies at $z \simeq 2$ matches that of
$\log_{10}(M_{\rm vir}/M_\odot) \simeq 12.7$ haloes in the same
epoch, and is substantially higher than what predicted by most
semi-analytic models for galaxy formation.

\section*{ACKNOWLEDGMENTS}
{Thanks are due to the referee for constructive comments, that
helped improving the paper.} Work supported in part by MIUR and
ASI.

\end{document}